\documentclass[a4paper,fleqn]{cas-sc}

\usepackage[numbers]{natbib}

\usepackage{lipsum}
\usepackage{graphicx}

\usepackage{lineno}

\def\tsc#1{\csdef{#1}{\textsc{\lowercase{#1}}\xspace}}
\tsc{WGM}
\tsc{QE}
\tsc{EP}
\tsc{PMS}
\tsc{BEC}
\tsc{DE}

\begin{document}
\let\WriteBookmarks\relax
\def\floatpagepagefraction{1}
\def\textpagefraction{.001}

\shorttitle{Impacts and Martian Shoreline Observability}

\shortauthors{MM Baum, SF Sholes, AD Hwang}

\title [mode = title]{Impact Craters and the Observability of Ancient Martian Shorelines}                      



%
\author[1]{Mark Baum}[orcid=0000-0001-8832-4963]

\cormark[1]


\ead{markbaum@g.harvard.edu}

\credit{Conceptualization, methodology, formal analysis, software, writing}

\affiliation[1]{organization={Harvard University Department of Earth and Planetary Sciences},
    city={Cambridge},
    state={MA},
    country={United States}}

\author[2]{Steven Sholes}[orcid=0000-0003-4854-1191]



\credit{Writing}

\affiliation[2]{organization={Jet Propulsion Laboratory, California Institute of Technology},
    city={Pasadena},
    state={CA},
    country={United States}}
    
\author[3]{Andrew Hwang}

\credit{Formal analysis}

\affiliation[3]{organization={College of the Holy Cross Department of Mathematics and Computer Science},
    city={Worcester},
    state={MA},
    country={United States}}

\cortext[cor1]{Corresponding author}



\begin{abstract}
The existence of possible early oceans in the northern hemisphere of Mars has been researched and debated for decades. The nature of the early martian climate is still somewhat mysterious, but evidence for one or more early oceans implies long-lasting periods of habitability. The primary evidence supporting early oceans is a set of proposed remnant shorelines circling large fractions of the planet. The features are thought to be older than 3.6~Ga and possibly as old as 4~Ga, which would make them some of the oldest large-scale features still identifiable on the surface of Mars. One question that has not been thoroughly addressed, however, is whether shorelines this old could survive modification and destruction processes like impact craters, tectonics, volcanism, and hydrology in recognizable form. Here we address one of these processes---impact cratering---in detail. We use standard crater counting age models to generate synthetic, global populations of craters and intersect them with hypothetical shorelines, tracking portions of the shoreline that are directly impacted. The oldest shorelines ($\geq$4~Ga) are at least 70~\% destroyed by direct impacts. Shorelines of any age $>$3.6~Ga are dissected into relatively short, discontinuous segments no larger than about 40~km when including the effects of craters larger than 100~m in radius. When craters smaller than 500~m in radius are excluded, surviving segment lengths can be as large as $\sim$1000~km. The oldest shorelines exhibit fractal structure after impacts, presenting as a discontinuous collection of lines over a range of scales. If the features are truly shorelines, high-resolution studies should find similar levels of destruction and discontinuity. However, our results indicate that observing shorelines as old as 4~Ga, should they exist, is a significant challenge and raises questions about prior mapping efforts.
\end{abstract}


\begin{highlights}

\item The oldest proposed shorelines ($\sim$4~Ga) would have been mostly destroyed by direct impacts.

\item Shorelines of any age >3.6~Ga would be dissected into relatively short, discontinuous segments shorter than 40~km.

\item Any putative shorelines should exhibit fractal segment lengths with a large number of gaps.

\end{highlights}


\maketitle

\section{Introduction}
\label{sec:introduction}
The presence of remnant shorelines on Mars, possible evidence of ancient oceans, has been debated since the Viking Orbiter returned images with high enough resolution to identify features on the scale of tens of meters \cite{parker_transitional_1989, baker_ancient_1991, parker_coastal_1993}. The proposed shorelines are the primary evidence for ancient oceans \cite{zuber2018}, and the debate surrounding them is important because the early martian climate remains somewhat mysterious \cite{wordsworth_climate_2016} and oceans imply long periods of relatively warm and habitable conditions on our neighboring planet. Confirming the presence of ancient oceans would constrain the early water budget, climate, and potential for habitability.

Several different shorelines have been proposed and investigated, but close inspection reveals problems with the interpretation of these features as shorelines \cite{dickeson_martian_2020}. First, the mapped features exhibit large variation in elevation, sometimes with large scatter over relatively small distances \cite{carr_oceans_2003, sholes_where_2021}. This is problematic because shorelines should be approximately level like the boundary of a water body, forming an equipotential surface due to gravity. Multiple global topographic deformation models have been proposed \cite{citron_timing_2018, perron_evidence_2007}, but none successfully account for all of the observed elevation variation \cite{SHOLES2022114934} and other geomorphological features thought to be genetically related to an ocean exhibit contradictory elevation patterns \cite{di_achille_ancient_2010, riverahernandez_deltas_2019, SHOLES2022114934}. Second, high-resolution studies of limited areas along the proposed shorelines find little or no evidence for the shoreline interpretation \cite{malin_oceans_1999, ghatan_paucity_2006, sholes_quantitative_2019, sholes_reassessing_2019}. Third, clear and usable location information for most of the proposed shoreline mappings has not been made publicly available. This lack of transparency has caused confusion and poor coordination. Only recently have proposed shoreline maps been digitized and the data made publicly available \cite{sholes_where_2021, sholes_steven_2020_3743911}.

In addition to issues related to the mapping and interpretation of the proposed shorelines, the warm and wet climate required to sustain a large ocean is difficult to justify. Independent lines of evidence suggest the cumulative time with warm surface environments on Mars was not more than $\sim$10$^7$~yr. The abundance of unaltered ancient igneous minerals, relative absence of carbonate rocks, and nature of many observed phyllosilicate sequences are consistent with a mostly cold early climate \cite{ehlmann_subsurface_2011, niles_geochemistry_2013, ehlmann_mineralogy_2014, bishop_surface_2018}. Geomorphic studies also conclude that less than $\sim$10$^7$~yr of active surface hydrology is required to erode the observed valley networks \cite{hoke_formation_2011}. Oceans during cold periods are also problematic, as climate models consistently indicate that without a yet unknown source of long-term warming, ocean water would have rapidly migrated to high-elevation and polar cold traps \cite{clifford_evolution_2001, wordsworth_global_2013, wordsworth_climate_2016, turbet_paradoxes_2019}. Recent modeling of stable northern oceans in cold climates \cite{schmidt2022circumpolar} does not fully account for the problem of southern highlands cold traps.

Although many different shorelines have been referenced \cite{parker_bookchapter_2010, parker_noachian_2020}, research has been focused on two primary proposed shorelines, the "Arabia Level" and "Deuteronilus Level" (the term "level" is used as a neutral, non-genetic descriptor). \citet{ivanov_topography_2017} found that the age of the Vastitas Borealis Formation (VBF), which largely demarks the Deuteronilus Level, clusters tightly around 3.6~Ga. \citet{citron_timing_2018}, applying topographic deformation models in an attempt to correct the putative shoreline elevation problems, inferred that the Arabia Level formed $\geq$4~Ga and the Deuteronilus Level formed 3.6~Ga.

A remnant shoreline as old as 4~Ga would be one of the oldest recognizable large-scale features on the surface of Mars. It would be roughly as old as the giant impact basins \cite{werner_early_2008} and older than the bulk of Tharsis \cite{anderson_primary_2001, citron_timing_2018}. The implications of such antiquity have not been thoroughly explored. In particular, it is natural to wonder whether such an old feature could survive to the present day in recognizable and mappable form. Active surface processes such as volcanism, impact cratering, hydrologic activity, and tectonics all have the potential to modify and/or destroy portions of possible shorelines after their formation. Aeolian activity, compounded over >3~Ga, could also substantially erode potential marine landforms. All of these processes would have been most intense early in martian history, so the preservation of an ancient shoreline may depend quite strongly on its proposed age.

Here we investigate the effect of just one destructive process on the preservation of ancient martian shorelines: impact craters. Crater populations have long been used for the dating of martian surface features, but crater counting statistics can also be applied in reverse. The age of a feature determines the specific population of craters that is expected to appear after its emplacement. We exploit this relationship by generating synthetic populations of craters with different ages and studying their aggregate effect on idealized representations of shorelines. In Section \ref{sec:simulations}, we explain the details of our synthetic crater populations and our simulations. In Section \ref{sec:results}, we detail the results of our simulations. Finally, in Section \ref{sec:discussion}, we discuss the implications of our results and consider the path forward.

\section{Simulations}
\label{sec:simulations}

\begin{figure}
	\centering
	\includegraphics[width=\textwidth]{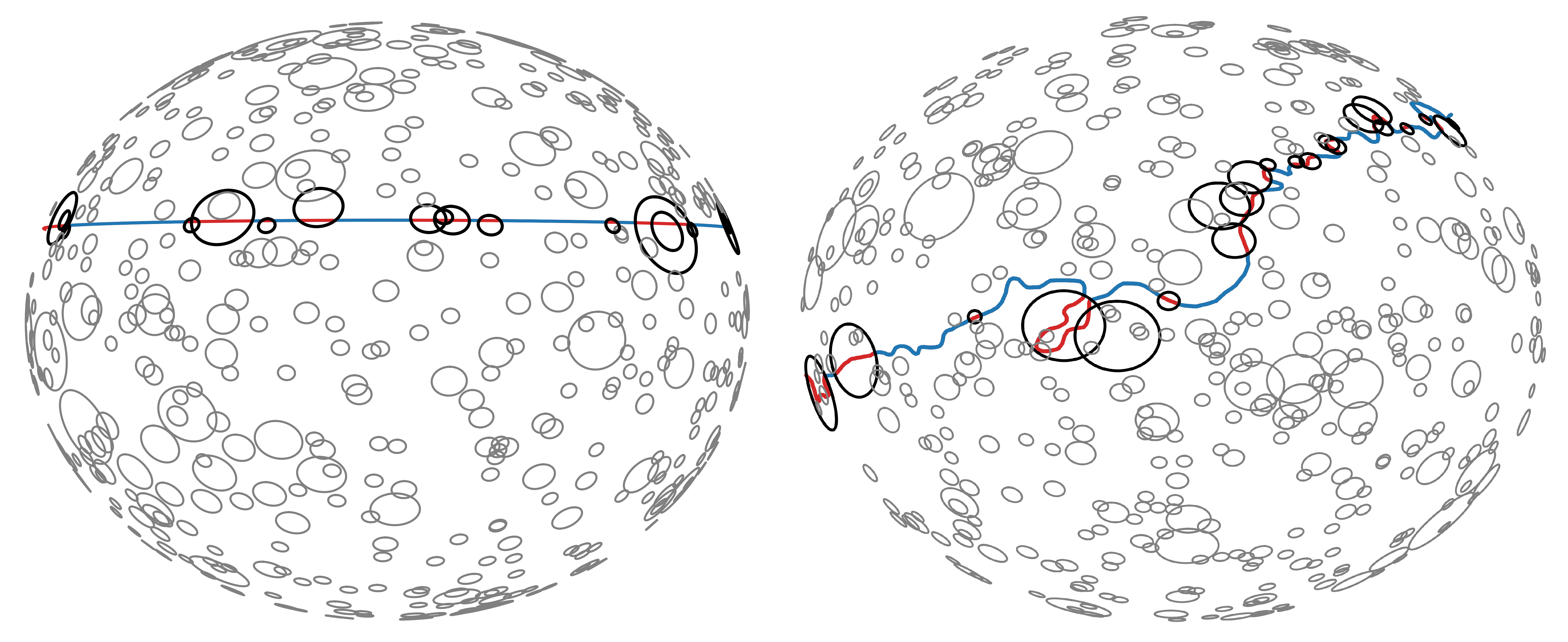}
	\caption{Orthographic representation of the simulations presented in this study. On the left, a uniform latitude putative shoreline (blue) is being intersected by various craters, with intersected portions highlighted in red. We refer to this as the "isolatitude" representation. Intersecting craters are drawn in black and all other craters are drawn in gray. On the right, the putative shoreline is represented by the coordinates of the proposed Arabia Level \cite{parker_coastal_1993, sholes_where_2021}. This is our "mapped" representation. In both of these examples, we only show a small number of artificially large craters for visual clarity.}
	\label{fig:examples}
\end{figure}

We generate synthetic, global populations of craters using the Hartmann \cite{hartmann_martian_2005} size-frequency bins, as detailed in \citet{michael_planetary_2013}. We choose to use synthetic crater populations, rather than a database of real craters because we think it is important to include the effects of craters that are smaller than those included in the available database \cite{robbins_new_2012, robbins_new_2012-1}. Table 1 in \citet{michael_planetary_2013} shows the crater density (units of km$^{-2}$Ga$^{-1}$) for bins with integer indices between -16 and 19. To generate a global crater population, we multiply each bin's density by the surface area of Mars ($1.44 \times 10^{14}$~m$^2$) and scale the result for population age using Equation 3 in \citet{michael_planetary_2013},
\begin{equation}
    \frac{e^{6.93 t} - 1}{1.5409 \times 10^{10}} + t \, ,
\end{equation}
where $t$ is the age of the crater population in Gyr. The resulting number of craters in each bin is rounded down to the nearest integer in this study. We checked our populations against \citet{palucis_quantitative_2020}, where a nearly identical sequence of calculations is made for smaller geographic domains.

Every crater within a bin is assigned the mean diameter for that bin,
\begin{equation}
    D_{\textrm{mean}}(i) = 2^{\left( i/2 + 1/4 \right)} \, ,
\end{equation}
where $i$ is the index of the bin. The location of each crater is assigned randomly over the surface of the spherical planet using
\begin{align}
    \theta &= \arccos \left( 1 - 2 X \right) \\
    \phi &= 2\pi X \, ,
\end{align}
where $X \in [0,1)$ is a uniformly distributed pseudorandom number drawn independently for each coordinate, $\theta \in [0,\pi)$ is the colatitude, and $\phi \in [0,2\pi)$ is the longitude. We use radians for the simulations but, where necessary, present our results using degrees.

These synthetic, global crater populations include a potentially enormous number of craters, increasing rapidly with age. For example, a 4~Ga population includes more than 2.3~billion craters larger than 100~m in radius. If the size and location of each crater were stored explicitly in computer memory as three 64 bit floating point numbers, this population would require about 55 GB of memory. Instead, we handle populations by computing and storing the number of craters in each bin along with a seeded random number generator (RNG). When iterating through a population, craters are assigned reproducible coordinates on the fly using the RNG. All of our crater population code is publicly available for inspection or reuse \cite{mark_baum_2022_5961048}.

\begin{figure}
	\centering
	\includegraphics[width=0.45\textwidth]{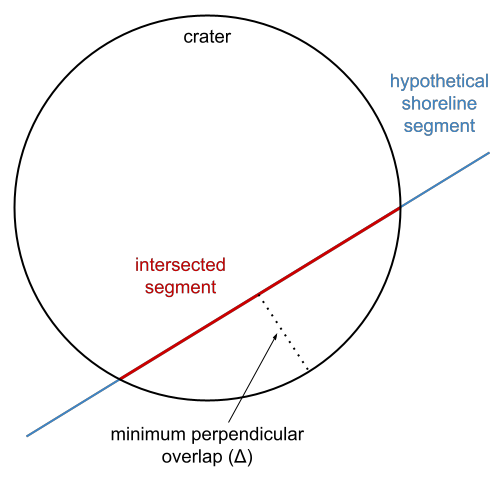}
	\caption{Diagram of a single crater intersecting a segment of hypothetical shoreline, illustrating the minimum perpendicular overlap ($\Delta$).}
	\label{fig:crater}
\end{figure}

Crater populations for different ages are then intersected with two different representations of the proposed shorelines, simulating the effect of impacts on shorelines of different ages. First, we represent a shoreline as a ring of constant latitude circling the planet, which we call the "isolatitude" representation. As we show later, this is a very good approximation for simulations with the real coordinates of the proposed shorelines, but the simple geometry enables rapid simulations and more accurate statistics. To check if a crater intersects this ring, we must only compare the crater's radius to the longitudinal distance between the crater center and the ring. When an intersection occurs, any portions of the hypothetical shoreline lying inside the crater are removed. An example of the isolatitude simulations is shown in the left portion of Figure \ref{fig:examples}.

Second, we intersect crater populations with an accurate representation of one proposed shoreline, the Arabia Level of \citet{parker_coastal_1993}, as reproduced by \citet{sholes_where_2021}. We represent this proposed shoreline with about 200 continuously connected geodesic segments and refer to it as the "mapped" representation. These segments are obtained by coarsening the original set of coordinates down to a target arc length of 0.05 radians (about 2.9{\textdegree} or 170~km on Mars). Simulations with this more accurate representation of the proposed shoreline are much more computationally expensive than the simple isolatitude representation because each crater must be tested for intersection with each of the shoreline segments. Further, testing intersections between arbitrary craters and geodesic segments is more involved than a simple comparison based on the longitude. However, just like the isolatitude simulations, whenever a crater intersects any of the mapped shoreline segments, portions of the segment lying inside the crater are removed. The right portion of Figure \ref{fig:examples} shows an example mapped simulation.

To understand the role of impact ejecta, we carry out simulations with an "ejecta multiple" between 1 and 2. This multiple directly scales the radius of all craters in a population. For example, given a crater with a radius of 1~km and an ejecta multiple of 1.5, we assume that ejecta would obscure the shoreline anywhere within 1.5 times the crater's radius. The simulation treats the original 1~km crater as a 1.5~km crater. An ejecta multiple of 1 represents the case where ejecta have no effect on obscuring or obliterating the putative shorelines and this baseline case is reported for all simulations.

For the isolatitude and mapped simulations, we collect statistics about the final state of the shoreline after impacts by crater populations representing ages between 4 and 3.6~Ga. Because the isolatitude case is much faster, we simulate 144 random realizations of the crater populations for each age. We choose 144 because our computing node has 48 processors, so it is most efficient to run trials in parallel batches that are multiples of 48. For the isolatitude cases, we also use a relatively dense sample of different ages, five different shoreline latitudes, and 11 ejecta multiples between 1 and 2. For the mapped Arabia Level, we simulate 8 realizations of crater populations for each age and ejecta multiples of 1, 1.5, and 2.

In all simulations, we exclude craters smaller than 100~m in radius. This is because of the dramatic computational cost of simulating increasingly small craters on the global scale and because the destructive effects of small craters on proposed shoreline features are less obvious. In all simulations, we also enforce a minimum perpendicular overlap ($\Delta$) that prevents marginal intersections from disrupting the hypothetical shoreline, as illustrated in Figure \ref{fig:crater}. We choose two values, 50~m and 500~m, which reflect how the results scale with ($\Delta$), and by extension mapping scale. For broad observable contacts, e.g., the Deuteronilus Level which follows the Vastitas Borealis Formation, larger mapping scales are sufficient with lower-resolution data (e.g., 100~m/px) and gaps are likely to only be included if they are $\gtrsim$1~km in length, so a $\Delta=500$~m is most relevant. For features that are more narrow, e.g., the proposed Arabia Level which exhibits a fairly narrow expression in high-resolution ($\sim$1~m/px) studies with widths on the order of 100~m, the lower value of $\Delta=50$~m is most relevant. Results are moderately sensitive to the choice of a 50~m intersection threshold, but a higher threshold near 100~m would not change our conclusions.


\section{Results}
\label{sec:results}

\begin{figure}
	\centering
	\includegraphics[width=\textwidth]{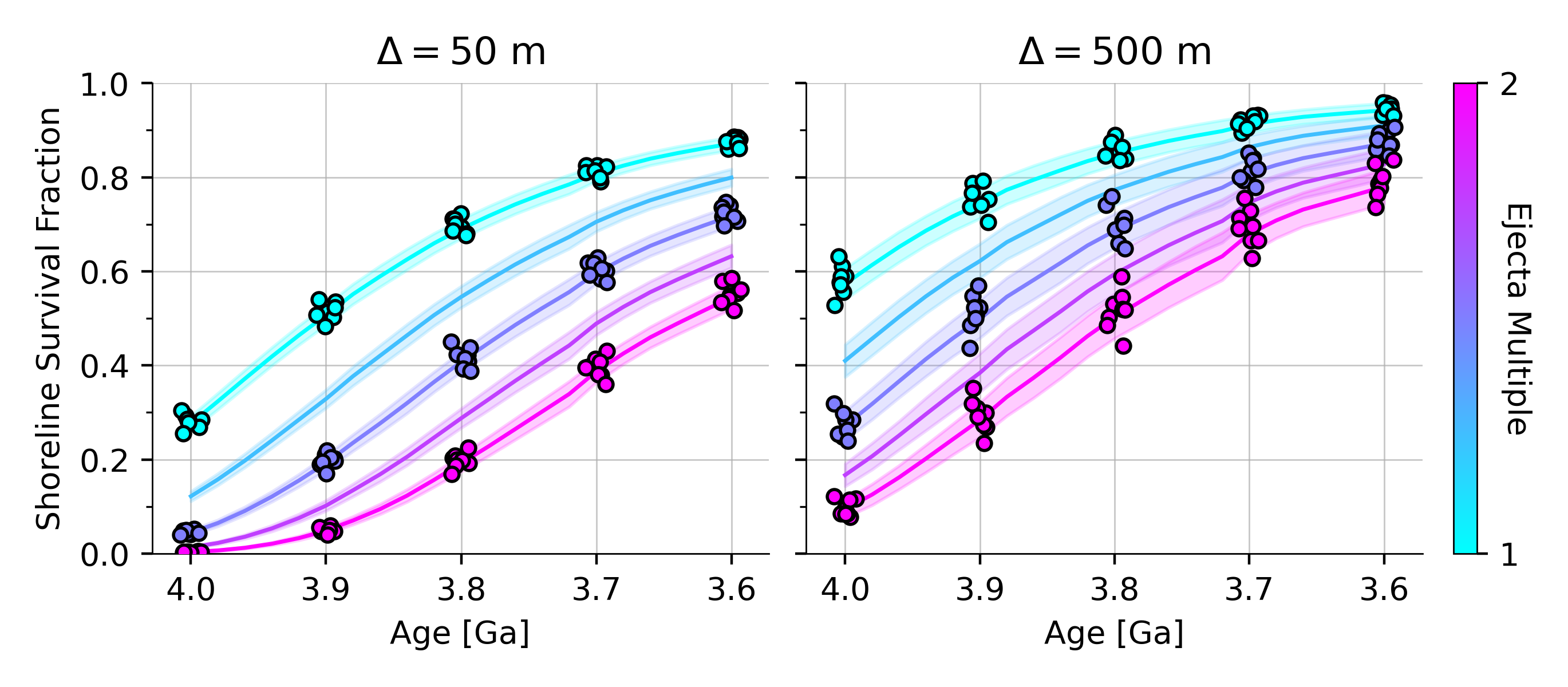}
	\caption{The fraction of simulated shoreline length remaining after impacts for different ages and ejecta multiples. Lines represent the mean result for isolatitude simulations ($\phi=30^{\circ}$) with the standard deviation shown in the accompanying bands. Each dot shows the result for one mapped simulation. Clusters of dots represent different realizations of synthetic crater populations. Each cluster of points has been spread out slightly in the horizontal dimension for visual clarity.}
	\label{fig:survival}
\end{figure}

Figure \ref{fig:survival} shows the fraction of a shoreline's original length that remains after simulation, for crater populations representing ages between 4 and 3.6~Ga, ejecta multiples between 1 and 2, and minimum perpendicular overlap ($\Delta$) of 50 and 500~m. Lines show the mean values of 144 isolatitude simulations with the shoreline at 30$^{\circ}$. Bands around each line indicate the standard deviation. Clusters of points on top of the lines represent the 8 simulations with each parameter combination for the mapped shoreline, where each point represents a single realization/result. We observe that mean survival fractions for the isolatitude (lines) and mapped (dots) simulations are practically indistinguishable.

The lightest blue lines in both panels of Figure \ref{fig:survival} show cases where crater ejecta play no role in obscuring the shoreline (ejecta multiple of 1). It represents the maximum fraction of a shoreline that would be expected to survive impacts. For $\Delta=50$~m, this line shows that at 4~Ga only roughly 30~\% of a hypothetical shoreline survives impacts. The survival fraction rises quickly for younger ages. At 3.9~Ga, a maximum of 50~\% is preserved, rising further to about 85~\% by 3.6~Ga. For $\Delta=500$~m, survival fractions are uniformly higher, with a maximum of 60~\% of a 4~Ga feature likely survive impacts.

The ejecta multiple strongly influences survival fractions. For example, with $\Delta=50$~m and an ejecta multiple around 1.5, where ejecta are assumed to obscure the shoreline anywhere within 1.5 times each crater's radius, less than 5~\% of any 4~Ga shoreline would be observable. With higher ejecta multiples, the area affected by craters saturates and the preserved fraction of a shoreline is nearly zero. The survival fraction is most sensitive to the ejecta multiple in the middle of our simulated ages. At 3.8~Ga, the difference between a multiple of 1 and a multiple of 2 is about 50~\%.

\begin{figure}
	\centering
	\includegraphics[width=\textwidth]{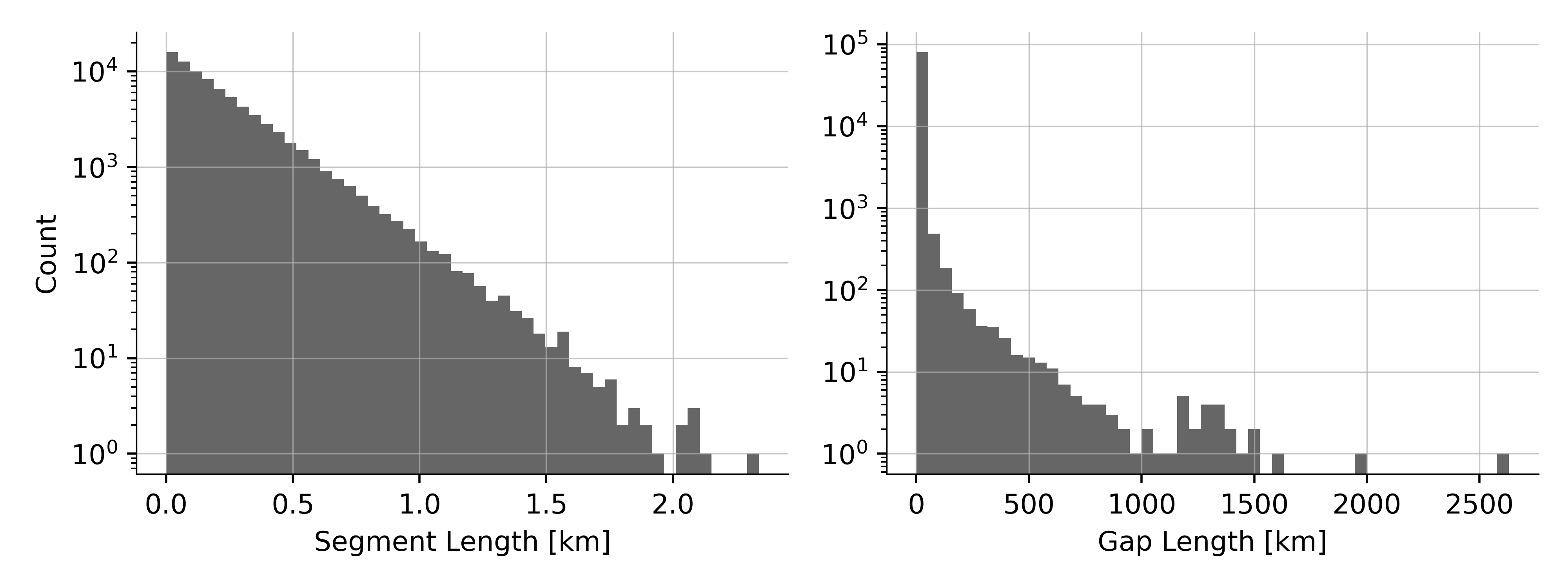}
	\caption{On the left, the distribution of remaining shoreline segment lengths after impacts. On the right, the distribution of gap lengths between the remaining segments. Both distributions were obtained from the combined results of 21 simulations with an isolatitude shoreline at 30$^{\circ}$, crater populations representing 4~Ga, and an ejecta multiple of 1.5.}
	\label{fig:histograms}
\end{figure}

Figure \ref{fig:histograms} shows the distribution of remaining shoreline segment lengths after simulation and the distribution of the gap lengths between segments. These distributions were obtained from the combined results of 21 simulations with an isolatitude shoreline at 30$^{\circ}$, crater populations representing 4~Ga, $\Delta=50$~m, and an ejecta multiple of 1. Note the logarithmic vertical axes.

The segment lengths are exponentially distributed. Because of this, long segments are extremely unlikely for these simulations at 4~Ga. The 50$^{\textrm{th}}$ percentile occurs at only 275~m and the largest segments are only about 5~km in length. The gaps between segments display an even stronger skew toward small lengths. This distribution is dominated by values below 100~km, a consequence of the much larger number of small craters in the population. The 50$^{\textrm{th}}$ percentile in gap length occurs at about 300~m. However, the probability of at least one large gap is not negligible. In this group of simulations, we observe several gaps larger than 500~km and one exceeding 1250~km, without including the effect of any ejecta.

\begin{figure}
	\centering
	\includegraphics[width=0.45\textwidth]{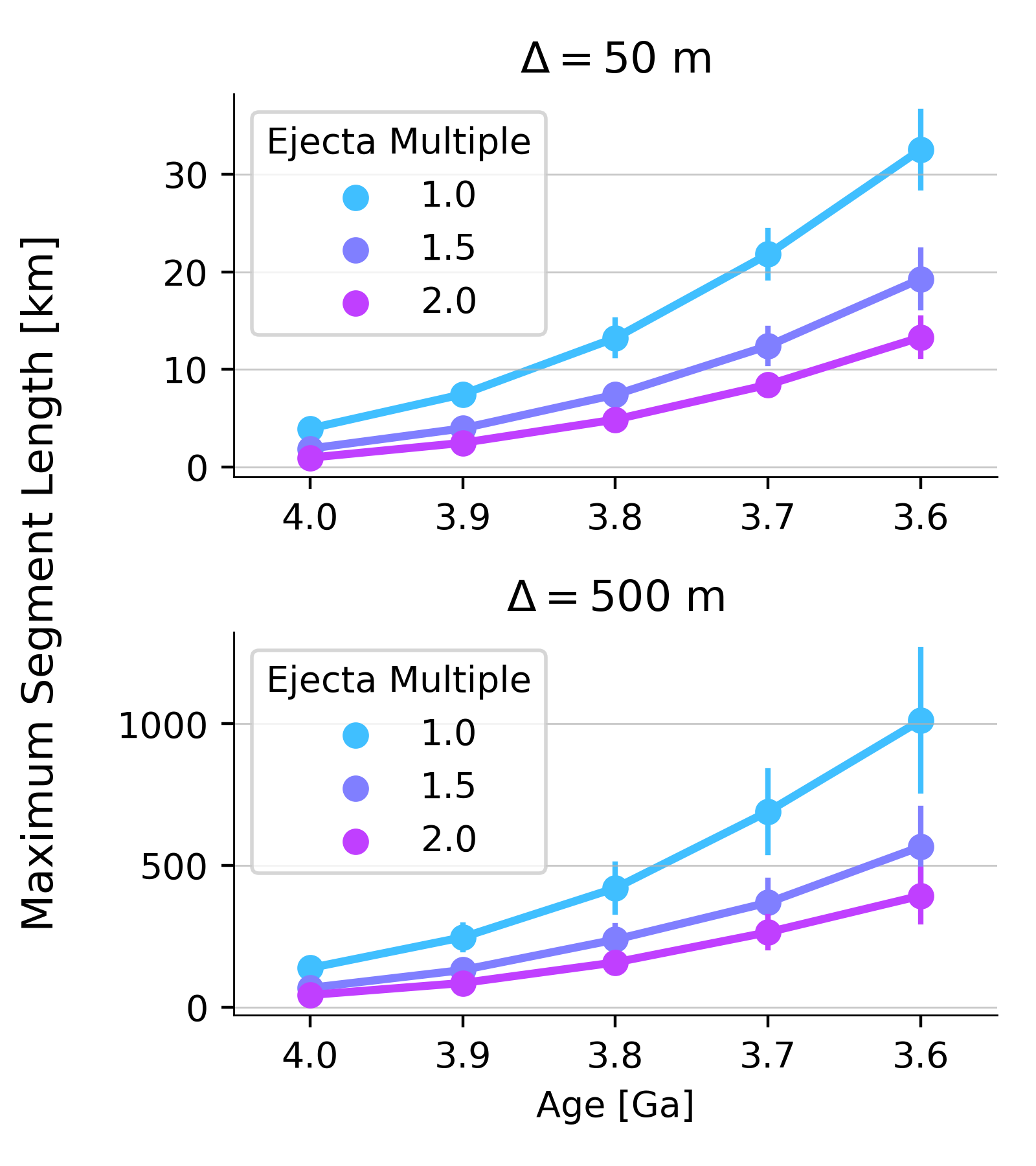}
	\caption{A summary of the maximum segment length for ages and ejecta multiples spanning our selected ranges. Each dot represents the average length of the largest segment or gap across 144 realizations. Vertical lines through the dots (where visible) represent the standard deviation of each group.}
	\label{fig:segs}
\end{figure}

Figure \ref{fig:segs} focuses on the largest segments after simulation, now including the results from all 144 realizations and different ejecta multiples. Each dot indicates the average length of the largest segment after simulation, with standard deviations indicated by the vertical bars. In agreement with Figure \ref{fig:histograms}, the top panel of Figure \ref{fig:segs} shows that continuous shoreline segments longer than about 5~km are unlikely for a shoreline age of 4~Ga and $\Delta=50$~m. For any age, however, the longest segment is very likely to be shorter than 40~km in this case. For $\Delta=500$~m, the longest surviving segments are dramatically longer because the total number of craters in each population is much lower.

\begin{figure}
	\centering
	\includegraphics[width=0.45\textwidth]{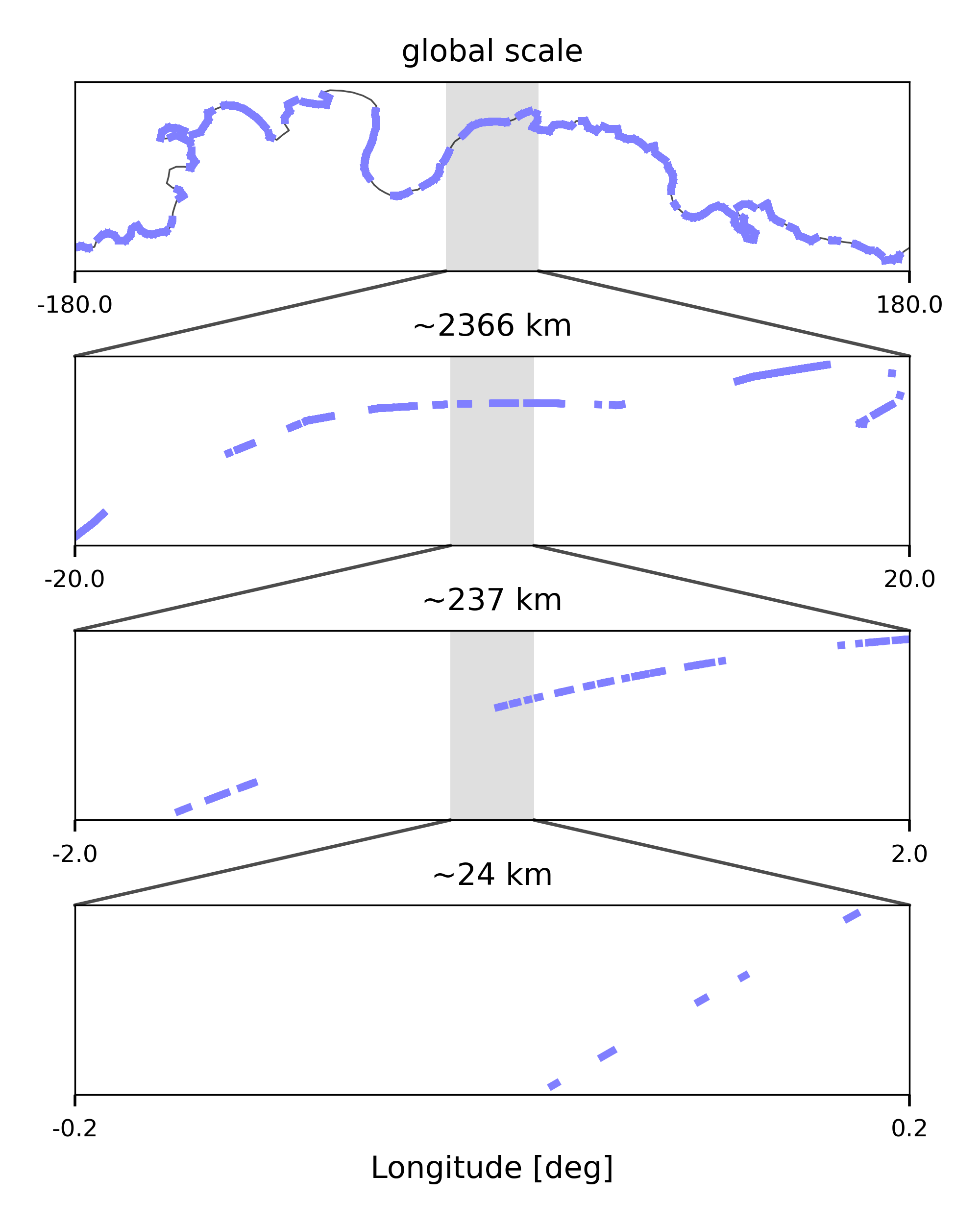}
	\caption{The fractal structure of shoreline segments after cratering using the mapped shoreline, an ejecta multiple of 1.5, and a crater population representing 4~Ga. The top panel shows the entire collection of final shoreline segments on the global scale. The original, intact shoreline is drawn in gray. Although the shoreline appears mostly intact on this scale, this is an artifact of the visualization and only about 2~\% of its original length is present. Lower panels zoom in to increasingly small scales, where the short, dissected segments become evident. In the lowest panel, where the horizontal scale is about 24~km and individual segments are finally visible, the shoreline is revealed to be mostly absent. Note that we zoom in to the exact center of each panel instead of choosing specific slices based on their appearance.}
	\label{fig:fractal}
\end{figure}

Finally, Figure \ref{fig:fractal} shows the final state of one 4~Ga simulation with an ejecta multiple of 1.5 and $\Delta=50$~m, across different spatial scales. At the global scale, only intersections with large craters are visible and small gaps aren't resolvable. Zooming in reveals the fractal structure of the final shoreline state. Shoreline segments that appear continuous on large scales are discontinuous on smaller scales, all the way down to a scale of about 40~km. At high enough resolution, the extent of the destruction is evident. The bottom panel of Figure \ref{fig:fractal} shows that the final shoreline is mostly composed of gaps between short segments.

\section{Discussion}
\label{sec:discussion}

Very old shorelines would be significantly disrupted by impact craters. Simulations indicate that less than 30~\% of the oldest proposed shorelines (4~Ga) would have survived direct intersection by craters when relatively small craters are included ($\Delta=50$~m). Accounting for burial by even a limited ejecta blanket decreases this percentage considerably. For example, an ejecta multiple of 1.3 pulls the survival fraction of a 4~Ga shoreline down to only $\sim$10~\%. For larger ejecta multiples, the original 4~Ga shoreline would be almost entirely obscured.

The age dependence of the shoreline survival fraction is striking (again, Figure \ref{fig:survival}). A narrow 4~Ga shoreline is mostly destroyed without any ejecta burial, but a 3.6~Ga shoreline is mostly preserved, even with our highest ejecta multiple (Figure \ref{fig:survival}). This result raises questions about the oldest proposed shorelines, most notably the Arabia Level. As mentioned in Section \ref{sec:introduction}, maps of the Arabia Level are inconsistent and imprecise \cite{sholes_where_2021}. Early studies present an Arabia Level that is continuous over large fractions of the planet's longitude \cite{parker_transitional_1989, parker_coastal_1993, clifford_evolution_2001} and later studies adopt coarse, partial reconstructions of these original studies \cite{carr_oceans_2003, perron_evidence_2007, citron_timing_2018}. There is no global consensus map of the Arabia Level, constructed from now globally-available high-resolution orbital data, to which we could meaningfully compare survival fractions and segment length statistics.

The shoreline survival fraction is clearly sensitive to the ejecta multiple and the role of ejecta is debatable in this context (again, Figure \ref{fig:survival}). We think it is reasonable to assume that any possible shoreline morphology inside crater rims would be unrecognizable after impact, but the long-term observability of shoreline morphology immediately outside a crater's rim is more uncertain. The effect of ejecta on the burial and observability of shoreline features is complex and depends on how well the ejecta blanket itself is preserved. It is common to observe continuous ejecta blankets extending 1-2 crater radii beyond the rims of fresh craters \cite{melosh1989impact}, equivalent to ejecta multiples of 2-3 in this study. However, the thickness of ejecta deposits decreases rapidly away from crater rims and small craters have much thinner ejecta blankets \cite{melosh_planetary_2011}. Small craters are more destructive than large ones in our study because of their high frequency, so a lower ejecta multiple is probably more appropriate in this context. This is why we simulate ejecta multiples less than two. Given the considerable uncertainty about the role of ejecta, however, we do not specify a preferred value and our main conclusions do not depend on ejecta multiples greater than one.

We only simulate impact craters down to radii of 100~m, but impact gardening by much smaller craters may be an important process to consider. Craters with radii as small as 0.25~m play an important role in turning the regolith and destroying narrow landforms (e.g., the putative shorelines) \cite{hartmann_martian_2001}. However, these small craters saturate the surface over time and are prohibitively expensive to simulate directly on the global scale. Future work could investigate the role of smaller craters in a limited geographic domain and the simulation code used in this study could be applied to such a project with very little modification. Other processes like hydrology, tectonics, and volcanism also have the potential to obscure old shorelines and could be studied in detail. Because all of these processes were most intense early in martian history, the likelihood of a 4~Ga shoreline surviving until the present with large portions intact seems quite low.

Finally, it is important to emphasize that the evidence for these proposed shorelines is currently lacking and detailed high-resolution observations are required to determine their validity, especially the Arabia Level. The location of the Arabia Level is so uncertain, and prior maps appear so unreliable, that future research should not adopt any one set of coordinates before the proposed feature has been carefully reexamined, preferably by more than one study. Fundamental questions remain unanswered, such as whether conditions on early Mars would be able to form erosional shorelines \cite{kraal2006, banfield2015}. As our results suggest, even if an ancient ocean existed on Mars, it may be unlikely that compelling evidence of this ocean can be assembled from orbital data. As Figure \ref{fig:fractal} suggests, because any old possible shorelines are likely to be so significantly obscured and segmented, large-scale surface features and patterns associated with a hypothetical shoreline may be better observational targets than the hypothetical shoreline itself.


\subsection*{Acknowledgements}
\label{sec:acknowledgements}
The computations in this paper were run on the FASRC Cannon cluster supported by the FAS Division of Science Research Computing Group at Harvard University. Computation was performed primarily using the Julia Language \cite{Bezanson2012, bezanson2017julia} with the DrWatson package \cite{Datseris2020} for reproducibility and organization. Figures were created using Seaborn \cite{Waskom2021} and Matplotlib \cite{Hunter:2007}. Code is publicly available at \texttt{github.com/markmbaum/shoreline-survival} and archived with Zenodo \cite{mark_baum_2022_5961048}. All other files required to reproduce this project are also archived with Zenodo \cite{mark_baum_2022_5961694}. Part of this research was carried out at the Jet Propulsion Laboratory, California Institute of Technology, under a contract with the National Aeronautics and Space Administration (80NM0018D0004).

\printcredits

\bibliographystyle{model1-num-names}

\bibliography{shoreline_survival}

\end{document}